\documentclass{aastex631}

\usepackage[version=4]{mhchem}
\usepackage{rotating}
\usepackage{hyperref}
\usepackage{comment}
\usepackage{float}
\usepackage[dvipsnames]{xcolor}
\usepackage[normalem]{ulem}

\usepackage{cancel}
\usepackage[normalem]{ulem}
\usepackage{soul}

\begin{document}
\newcommand{\glm}[1]{\textcolor{ForestGreen}{#1}}
\newcommand{\req}{R_{\rm eq}}
\newcommand{\bmat}[1]{\mathrm{\textbf{#1}}}
\newcommand{\bvec}[1]{\textit{\textbf{#1}}}
\newcommand{\de}{\mathrm{d}}

\title{The T-matrix method for modeling the optical properties of grains in exoplanetary atmospheres.}

\author[0000-0003-3683-0834]{G. Aresu} 
\affiliation{INAF - Astronomical Observatory of Cagliari, Via della Scienza 5, Selargius (CA) (Italy)}

\author[0000-0001-8553-499X]{G. La Mura}
\affiliation{INAF - Astronomical Observatory of Cagliari, Via della Scienza 5, Selargius (CA) (Italy)}

\author[0000-0003-0602-6669]{G. Mulas}
\affiliation{INAF - Astronomical Observatory of Cagliari, Via della Scienza 5, Selargius (CA) (Italy)}

\author[0000-0001-7480-0324]{C. Cecchi-Pestellini}
\affiliation{INAF - Astronomical Observatory of Palermo, P.zza del Parlamento 1, Palermo (PA) (Italy)}

\author[0000-0003-0467-8049]{M. A. Iat\`{i}}
\affiliation{CNR-IPCF, Institute for Chemical and Physical Processes, Viale Stagno D'Alcontres 37, Messina (ME) (Italy)}

\author[0009-0007-0916-9469]{S. Rezaei}
\affiliation{CNR-IPCF, Institute for Chemical and Physical Processes, Viale Stagno D'Alcontres 37, Messina (ME) (Italy)}

\author[0000-0002-9882-1020]{A. Petralia}
\affiliation{INAF - Astronomical Observatory of Palermo, P.zza del Parlamento 1, Palermo (PA) (Italy)}

\author[0000-0002-9824-2336]{D. Locci}
\affiliation{INAF - Astronomical Observatory of Palermo, P.zza del Parlamento 1, Palermo (PA) (Italy)}

\author[0000-0002-9900-4751]{G. Micela}
\affiliation{INAF - Astronomical Observatory of Palermo, P.zza del Parlamento 1, Palermo (PA) (Italy)}

\author[0000-0002-5970-9001]{F. Saija}
\affiliation{CNR-IPCF, Institute for Chemical and Physical Processes, Viale Stagno D'Alcontres 37, Messina (ME) (Italy)}

\author[0000-0002-5823-9749]{R. Saija}
\affiliation{Department of Mathematics and Computer Sciences, Physical Sciences and Earth Sciences - University of Messina, Viale Stagno D'Alcontres 31, Messina (ME) (Italy)}

\begin{abstract}
Observations of exoplanetary atmospheres have revealed a rather complex co-existence between the gaseous and condensed phases, ranging from atmospheres apparently covered in clouds to clear ones, across a myriad of intermediate regimes. The interaction between solids and stellar radiation is of fundamental importance when forward or retrieval models are compared to observations. The simplest approach is to assume spherical particles, whose scattering and absorption properties are readily computed with Mie theory. More advanced computational methods, however, can also treat non-spherical condensates. Among these, the transition matrix approach has proven to provide a rigorous and efficient calculation of the extinction by non-spherical particles over a broad range of sizes and refractive indices. We here focus on the absorption and scattering cross sections of forsterite, spinel and silicon monoxide from the near to the far infrared range for particles of different sizes, with shapes ranging from compact ensembles of spheres to porous aggregates. We find that, depending on the size and the configuration chosen, the differences in extinction can be as high as a factor of $2-3$, with structured grains being more efficient absorbers in the near infrared for sizes greater than 1 $\mu$m. In a fiducial transmission spectrum we show that in the $\lambda = 1-5~\mu$m range the amount of spherical material needed to match the contribution of the structured grains can be a factor 2 higher. Such calculations, available for the community through a Zenodo repository, prove to be useful for interpreting recent and future observations of exoplanetary atmospheres.
\end{abstract}

\section{Introduction} \label{intro}
Recent years have seen remarkable progress in characterizing exoplanet atmospheres, particularly through the detection of major volatile and metal-bearing species such as \ce{CO2}, \ce{H2O}, \ce{CO}, and metal oxides, primarily in the highly irradiated atmospheres of hot Jupiters, warm Neptunes, and temperate gas giants \citep{Tinetti07, deKok13, Barman15, Evans16}. However, these early observations often encountered an observational challenge: flat, muted, or even featureless transmission spectra caused by high-altitude atmospheric aerosols, generally classified as condensation clouds and photochemically produced hazes \citep{Knutson14, Sing16, Grant23}. GJ 1214b became a reference target for these issues, as high-precision transmission spectroscopy yielded a remarkably flat spectrum across the optical and near-infrared regimes \citep{Kreidberg14}. Recently, JWST observations have begun to break this observational barrier, revealing key water vapor absorption features in emission and pointing to a highly reflective, metal-rich atmosphere with a high Bond albedo \citep{Kempton23}. Furthermore, near-infrared transmission spectroscopy has successfully detected signatures of \ce{CO2} and \ce{CH4}, suggesting a heavily enriched, super-solar metallicity. JWST observations have also revealed photochemically produced \ce{SO_2} in the hot Saturn WASP-39b \citep{Tsai23}  and large-scale, weather-like temporal variations in WASP-121b \citep{Changeat24}. These observations have begun to resolve the spatial and temporal dynamics of exoplanetary atmospheres, establishing that clouds are not static, uniform decks but active components of global meteorological cycles. Recent transit observations of the tidally locked hot Jupiter WASP-94A~b have revealed that the cooler morning limb is heavily blanketed by high-altitude magnesium silicate \ce{MgSiO3} clouds that obscure gaseous signatures, while the hotter evening limb is clear and displays strong water vapor absorption \citep{Mukherjee26}. This represents the first direct, spatially resolved detection of a planet-wide cloud cycle in action on a hot Jupiter. Additionally, JWST has provided further direct evidence for exotic aerosol compositions across cooler temperature regimes, such as salt clouds (e.g., \ce{KCl} and \ce{ZnS}) on the cold planet GJ 504b, which cover the atmosphere's deeper layers and weaken molecular features \citep{Baburaj26}.

The physical nature of these atmospheric particles, including their size, shape, composition, and spatial distribution, depends on local temperature and pressure conditions, as well as on radiative and dynamical processes. Condensation clouds form when gas-phase species become supersaturated along the local atmospheric pressure–temperature profile, nucleating and condensing into solid or liquid particles \citep{Gao21}. 

In particular, the general consensus is that solid condensation initially proceeds via nucleation, followed by grain growth through coagulation. These condensates naturally assemble into porous, fluffy aggregates whose reduced effective density and high surface-to-volume ratio enhance vertical lofting and mixing compared to compact, sedimentation-prone spheres \citep[e.g.,][]{Marley13, Ohno20, Vahidinia24}.

While traditional atmospheric modeling has relied on simplified grey cloud decks or assumed compact, homogeneous spheres to directly apply classical Mie theory, this spherical approximation fundamentally neglects the structural complexity of real atmospheric particles. Extensive remote sensing in the Solar System, climate studies, and laboratory analog experiments have demonstrated that aerosols naturally form highly irregular, crystalline, and fluffy aggregate geometries (e.g., \citealt{Lodge24}). 

The structural complexity of aggregates has a well-documented impact on their optical properties. For example, \citet{Petrova2000} demonstrated that cometary dust exhibits enhanced absorption at larger size parameters when modeled as aggregates rather than spheres. A similar morphological dependence was highlighted by \citet{Silsbee16,Shen2008}, who showed that compact spheres experience a sharp drop in relative extinction efficiency for sizes exceeding $\sim 0.7\,\mu$m; at smaller sizes, irregular grains maintain a more persistent extinction cross-section per unit mass.

The effect on the transmission spectrum of aggregates as opposed to spherical particles has been investigated by various authors focusing on the photochemical hazes on GJ 1214b (see e.g. \citeauthor{Adams19}, \citeyear{Adams19} and \citeauthor{Ohno20}, \citeyear{Ohno20}) finding that structured particles composed of spherical monomers are more likely to cause its dominant grey absorption, obscuring molecular features in the optical and infrared range.

Consequently, transitioning from basic parametric cloud models to physically consistent, non-spherical aerosol frameworks is a prerequisite for exoplanetary atmospheric studies. 

Accurate calculations of aerosol radiative transfer require solving for the extinction efficiency, single-scattering albedo, and scattering asymmetry parameter. In this work, we exploit the transition matrix (T-matrix) approach developed by \citet{Borghese07} as implemented in the open-source, nano-particle transition matrix code by \citet{LaMura26}, which provides a rigorous, highly stable framework to compute the scattering and extinction properties of realistic, arbitrary particles represented as assemblies of spherically symmetric subparticles with known optical constants. While classical T-matrix formulations can experience numerical instabilities for particles with large size parameters or extreme aspect ratios \citep{Waterman71}, representation via spherical sub-units allows for precise calculations of complex, irregular geometries. Moreover, the T-matrix approach yields exact analytical solutions for the orientational averages over all random orientations \citep{Borghese80}. Conversely, other methods, such as the Discrete Dipole Approximation (DDA) \citep{Draine94}, require a complete and costly rerun of the numerical solver to describe new orientations between the particle and the incident electromagnetic fields. To bypass the computational demand of exact methods, other studies in exoplanet literature have adopted the Modified Mean Field (MMF) theory developed by \citet{Tazaki18}. MMF is an analytical approximation that simplifies aggregate scattering and absorption by treating the constituent monomers as a continuous, statistically distributed mean-field medium, thereby accounting for internal multiple scattering and radiative coupling between monomers at low computational cost. Although MMF is highly efficient, it remains an approximation; exact T-matrix calculations provide arbitrary accuracy by fully capturing wave-optical coherent interactions and tracking individual scattering events between particles in the aggregate with all multiple-scattering sequences. 

In this work we build absorption cross sections for three condensates (forsterite, spinel and silicon monoxide) by assembling their structure as a collection of spherical monomers forming different arrangements: compact, elongated and porous. To quantify the impact of different structures on the transmission spectrum of a representative hot Jupiter, we exploit the equilibrium condensation module (based on \citealt{Kitzmann24}) within our 1D chemical code \texttt{cp\_zephyr}. The code can also account for the cloud structure of a given condensate solving for the balance between vertical mixing and sedimentation as detailed in \citet{Ackerman2001} and \citet{Batalha25}.  This comprehensive approach provides a self-consistent framework for linking atmospheric chemistry, non-spherical condensate microphysics, and spectral observables, offering insight into how clouds and disequilibrium processes influence the spectral appearance of transiting hot Jupiters. Our work is organized as follows: in Section \ref{P1} we outline our setup for the T-matrix calculation, in Section \ref{P3} we describe how structured particles are assembled while in Section \ref{P4} we compare the calculated cross sections against the ones used for simple spherical particles. 
Section \ref{P5} is concerned with the setup chosen to represent our fiducial atmospheric model, while Section \ref{P6} describes the calculation of the transmission spectra. In Section \ref{P7} we present our final remarks.

\section{On the viability and performance of the T-matrix method}\label{P1}

In the recent literature the calculations for the condensate opacities in many forward or retrieval models of exoplanetary atmospheres are often based on Mie theory under the assumption of spherical particles, see for example \citet{MacDon23} and \citet{Mullens24} with \texttt{POSEIDON}, \citet{Molliere19} and \citet{Carone26} with \texttt{PetitRadTrans}, \citet{Voyer26} with \texttt{TauRex} or other packages such as Photochem \citep{Wogan24} or even GCM models \citep{Mehta26}.

The processes that shape solid condensation in these environments are nevertheless believed to favour the formation of fluffy aggregates, whose dynamical properties, together with their response to the stellar radiation, can change depending on their aggregation state. Furthermore, this feeds back into the local dynamics as fluffier particles are easier to lift up to the low-pressure layers, ultimately impacting on the planet's transmission spectra \citep{Vahidinia24}.

Structured aggregates can be modeled using various theoretical approximations. The superposition T-matrix method applied here assumes that particles are assemblies of smaller spherical monomers, providing a versatile framework to explore parameter space in terms of overall particle morphology, monomer counts, and individual grain sizes. In principle, monomer sizes and chemical compositions can vary heterogeneously within a single aggregate, allowing complex, multi-component structures to be still solved within the same exact wave-scattering formalism.

To model these complex geometries, the superposition T-matrix method analytically solves Maxwell's equations for the multi-sphere assembly using vector spherical wave functions, computing orientation-averaged optical properties in a single step without the need to re-evaluate the system over discrete orientation angles \citep{Borghese07}.

To build the extinction cross-sections we used \texttt{NP\_TMcode} v0.10.5, whose implementation and historical development are detailed in \citet{LaMura26}\footnote{https://github.com/ICSC-Spoke3/NP\_TMcode}. The code originates from legacy \texttt{Fortran 66} code \citep{Saija01,Borghese07} and has been recently rewritten in C++, where it takes advantage of high performance parallel hardware architectures, substantially reducing the computing time, even for increasingly complex models.

We validate this approach by comparing our extinction cross-sections against corresponding calculations that reproduce the same compact aggregate structure with an equivalent radius (defined below) of $3\, \mu$m using the \texttt{ADDA} implementation of DDA \citep{Yurkin2011}. As illustrated in Fig.~\ref{abs_sca} for the extinction cross section of a compact aggregate (solid black line vs. diamonds), the agreement between the two methods is excellent. The mean relative difference between our T-matrix implementation and \texttt{ADDA} is $\sim 1\%$ for forsterite, $\sim 1\%$ for silicon monoxide, and $\sim 1.5\%$ for spinel up to $16\,\mu\text{m}$.

The notable discrepancy for spinel beyond $16\,\mu\text{m}$ stems from the behavior of the real part of its refractive index within the Reststrahlen band, where $n \ll 1$ and the material becomes highly reflective (metal-like). This induces strong localized surface resonances that volume-discretization methods like DDA struggle to fully resolve without computationally prohibitive dipole grid resolutions. On the other hand it should be noted that the T-matrix formalism, due to its reliance on spherical multipole expansions, can experience numerical difficulties when modeling structures with perfectly sharp edges. Nevertheless, we do not expect this to be a significant factor here, given that aerosol nucleation and growth in exoplanetary environments generally produce amorphous, irregular, or fluffy aggregates.

On equivalent cluster hardware, running \texttt{ADDA} in its standard multi-threaded CPU configuration (32 cores) required 6.25 hours for about 40 wavelength points, whereas our GPU-accelerated \texttt{NP\_TMcode} completed the calculation in just 10 minutes. This speedup is driven both by GPU parallelization and by the T-matrix method's ability to evaluate orientational averages analytically rather than through iterative angular sampling.

Overall, the close agreement observed between different numerical approaches, when used to simulate similar structures, validates our T-matrix calculations and confirms their reliability in computing exact extinction cross-sections for the aggregate materials considered in this work.

\section{Condensate Amorphous Structure and calculations}\label{P3}

\begin{figure}
    \centering
    \includegraphics[scale=0.4]{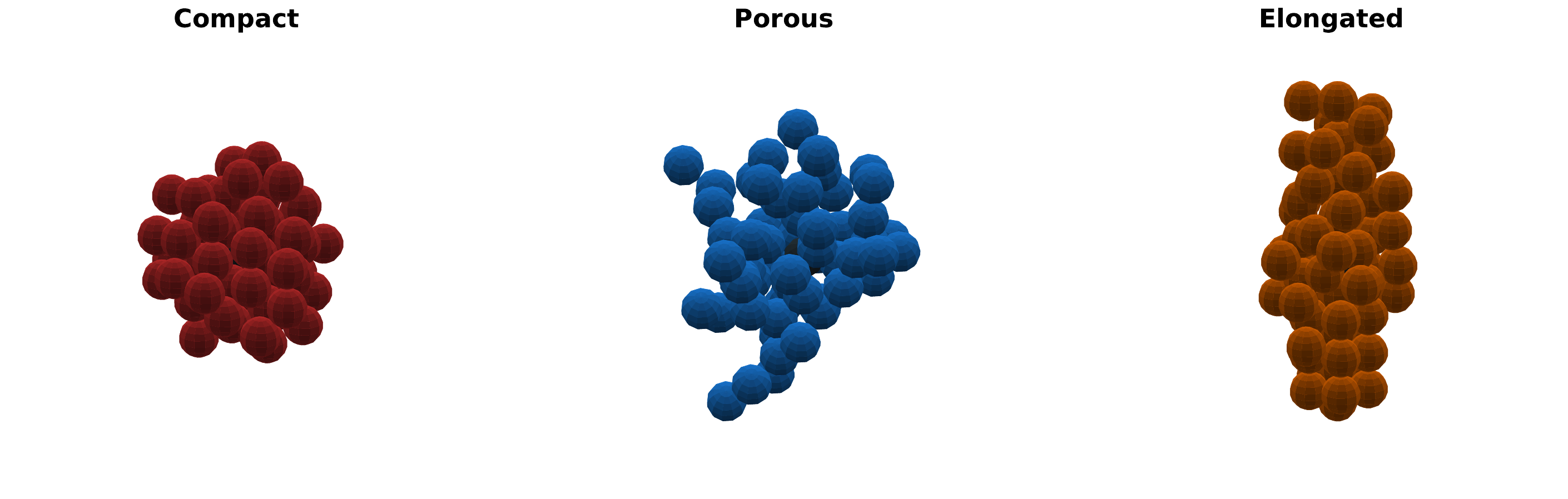}
    \caption{An example of the three different modes chosen for the agglomerate structure, the equivalent particle radius is 0.8 microns while the number of monomers is 64.}
    \label{agg_struct}
\end{figure}

To compute scattering, absorption and extinction cross sections for the solid phase of forsterite (\ce{Mg2SiO4}), spinel (\ce{MgAl2O4}) and silicon monoxide (\ce{SiO}) we assume different modes for the condensate agglomeration: \texttt{single sphere}, \texttt{compact}, \texttt{porous} and \texttt{elongated} (Fig. \ref{agg_struct}). The \texttt{single sphere} is solved using Mie theory and serves as a reference, for the other cases we build agglomerates of a given size out of smaller monomers (with radius $R_{\rm mon}$) until the equivalent radius of the whole particle $\req$ is equal to the radius of the sphere having the same mass:
\begin{equation}
    R_{\rm eq} = N^{1/3}R_{\rm mon}
\end{equation}
where $N$ is the number of spheres with that radius. 

We construct aggregates consisting of N spherical monomers of radius $R_{\rm mon}$ and compute their optical properties, comparing them with those of an equal-mass homogeneous sphere with equivalent radius $R_{\rm eq}$. The aggregate population covers equivalent radii from 0.1 to 1.0 $\mu\mathrm{m}$ in steps of 0.1 $\mu\mathrm{m}$, and from 1.0 to 10.0 $\mu\mathrm{m}$ in steps of 0.5 $\mu\mathrm{m}$. To keep the calculations computationally feasible, whenever an aggregate would contain more than 80 monomers, the monomer radius is increased for the next size step so that the total number of monomers remains below this limit. In addition, for aggregates with $R_{\rm eq} > 5 \mu\mathrm{m}$, the number of monomers is further restricted to fewer than 20 to limit the computational cost of the calculations.

In the \texttt{compact} case, we build upon the first sphere accommodating each new sphere in such a way that the distance is closest to the origin while physically touching a pre-existing sphere. In the \texttt{porous} mode, a new sphere is launched from a random direction toward the origin until it hits the pre-existing structure, thereby causing a more sparse configuration. The \texttt{elongated} mode is similar to the compact one, but the $z$ direction is preferred over the other two, with a ratio of 2.5:1:1. See figure \ref{agg_struct} for a selected sample and table \ref{size_table} in the appendix for the values used to build the agglomerates in terms of the number of spherical monomers with their radii, equivalent mass sphere and radius of the smallest sphere encircling the whole aggregate. The extinction cross section for each particle is then calculated in a wavelength interval ranging from 1 $\mu$m to 25 $\mu$m with steps of 0.1 $\mu$m. 

The T-matrix method implies a formally infinite multipole field expansion, which for practical calculations must be truncated at an order high enough to ensure accurate solutions, but low enough to avoid numerical instabilities, as including progressively smaller terms in the matrix might cause exceedingly high values in the subsequent matrix inversion. As detailed in \cite{LaMura26}, when applying the formalism to an aggregate of particles, we must choose a value for the internal field expansion ($l_i$, related to the largest monomer size) and a value for the external field expansion ($l_e$, related to the aggregate size). In general, these values need to be higher when the ratio of the monomer and the aggregate size with respect to the radiation wavelength becomes larger.

We determine the optimal expansion orders, $l_i$ and $l_e$, through a systematic convergence analysis. For each material, we first simulate the extinction cross-section of a single sphere across the range of sizes listed above, using, as reference, the shortest wavelength explored by our study ($\lambda = 1\, \mu\text{m}$).

The expansion order $l$ is 
progressively increased until the relative variation in the extinction cross section is lower than 10$^{-4}$. For an aggregate of equivalent radius $R_{\text{eq}}$ composed of monomers with radius $R_{\text{mon}}$, we set $l_i$ based on the convergence requirement of the individual monomer while $l_e$ accounts for the sphere encircling the entire cluster. As the calculation proceeds into longer wavelengths, the code automatically lowers the maximum expansion orders, to prevent the injection of numeric instabilities, whenever it detects that the ratio between the particle size and the radiation wavelength has dropped enough to still satisfy Wiscombe's numeric convergence criterion with an order reduced by one \citep{Wiscombe80}.

To calculate the optical cross-sections, the complex refractive indices $m = n + ik$ of the constituent materials must be specified. The selection of the optical constants for the condensates considered here depends on the availability and spectral coverage of existing laboratory data and computational simulations. For both forsterite and spinel, the data are obtained from a single, self-consistent literature source. For \ce{SiO}, however, no single reference spans the entire wavelength range of interest. Consequently, a composite set was constructed by combining data from multiple literature sources. We use amorphous forsterite, whose complex refractive indices are taken from \citet{Jager03}. This dataset covers an extensive spectral range from 0.2 to 500~$\mu$m, spanning both the stellar scattering and planetary thermal emission regimes. The optical constants were derived from pure, iron-free magnesium silicate samples synthesized via the sol-gel method, with the final indices calculated using a combination of Kramers-Kronig analysis and Lorentz-oscillator fitting of laboratory reflectance and transmission measurements. We also performed calculations using crystalline forsterite optical constants \citep{Zeidler11}. However, for the particle sizes explored in this work, the differences between the predicted extinction cross sections for spherical and crystalline forsterite are qualitatively similar to the ones obtained with amorphous forsterite. Consequently, the results presented here are based exclusively on the amorphous dataset. For the spinel, we use the complex refractive indices derived by \citet{Fabian01} from reflectance spectra using the classical Lorentz oscillator model. Finally, we exploit a composite set of optical constants for \ce{SiO} constructed by merging measurements from several laboratory sources. The optical and near-infrared refractive indices, $0.2 < \lambda < 2~\mu$m, are taken from the classic thin-film measurements of \citet{Hass54} and \citet{Philipp71}, obtained from optical reflectance and transmission experiments. The same dataset of \citet{Hass54} is adopted over $2 < \lambda < 5~\mu$m, while the thermal infrared, $\lambda > 5~\mu$m is described using the recent optical constants of \citet{Herguedas23}, derived from Fourier-transform infrared spectroscopy of \ce{SiO} thin films.

\section{Results}\label{P4}
\begin{figure}
    \centering
    \includegraphics[scale=0.8, trim=2.0cm 0mm 0mm 1cm, clip]{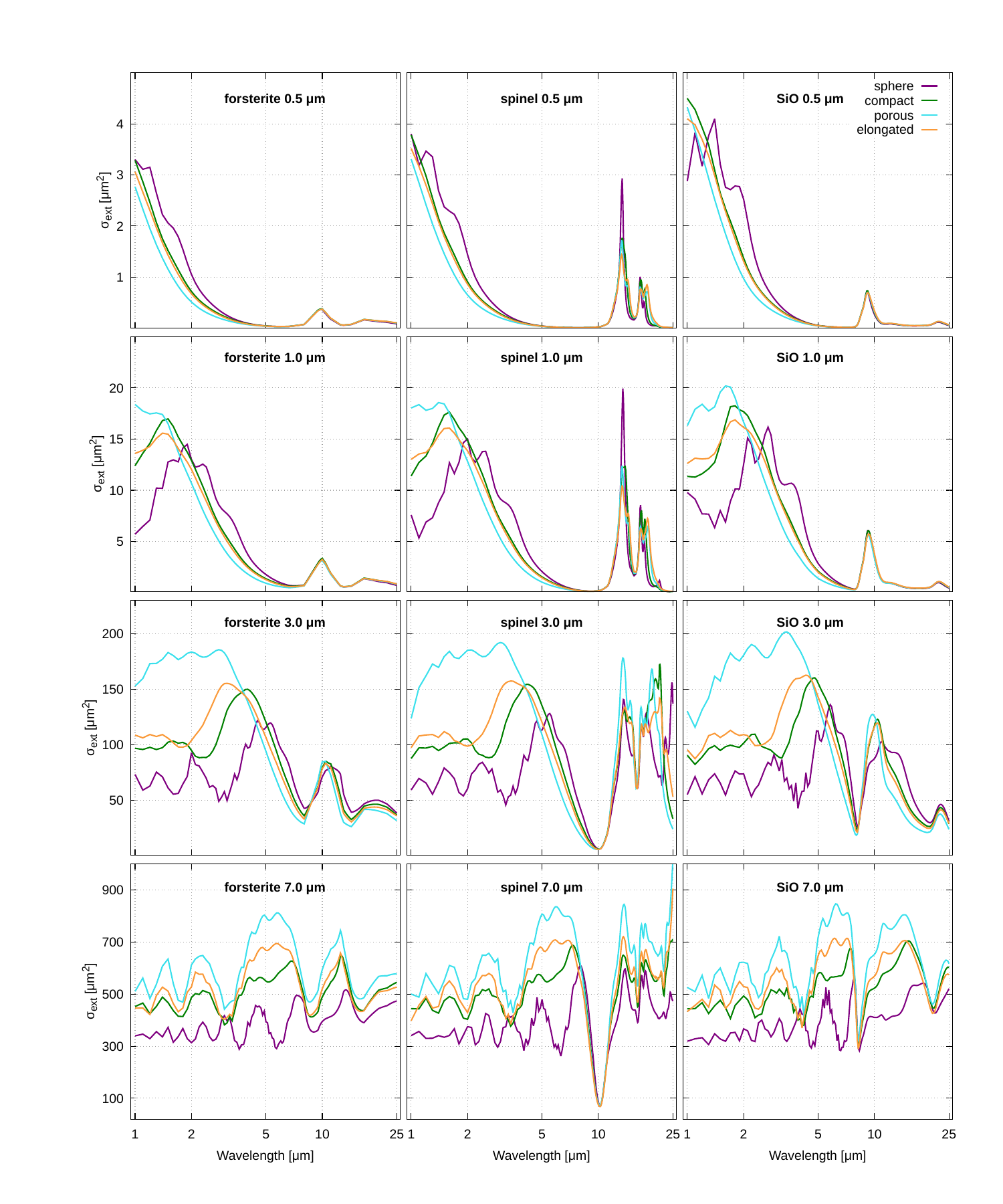}
    \caption{A sample of the extinction cross sections calculated for forsterite (left), spinel (middle) and silicon monoxide (right) in the range 1-25 micrometers using the $\texttt{NP\_TMcode}$. Moving down the particle size grows from 0.5 to 7.0$\mu$m. Color coded is the morphology chosen: single sphere (purple), compact (green), porous (cyan) and elongated (orange).}
    \label{cs_res}
\end{figure}

In Figure \ref{cs_res}, we present the calculated extinction cross-sections,  analytically averaged over random particle orientations, for a representative subset of particle equivalent radii across the three materials under investigation. A key finding of this study is that as the grain equivalent radius exceeds 1 $\mu$m, the extinction cross-section of the complex morphologies is consistently higher than that of the spherical case within the 1–5 $\mu$m wavelength range. 

For all materials as the equivalent radius of the agglomerate grows beyond a few micrometers, we find higher extinction at all wavelengths (see for example the bottom panels in figure \ref{cs_res}). Moreover, the compact spherical structure is systematically less absorbing than the elongated or porous mode. The latter seems to be the most effective structure in terms of extinction properties.

At sub-micron equivalent radii ($< 1$ $\mu$m) for forsterite and spinel particles, the spherical model is only marginally more efficient than the non-spherical structures. This is slightly less pronounced for SiO as the cross sections overlap around 1--2 $\mu$m. However, as $\req$ increases, the non-spherical grains become significantly more opaque to radiation at wavelengths between 1 and 5 $\mu$m. Beyond this range, the differences between the sphere and the other morphologies become negligible. Once $\req$ of approximately 5 $\mu$m is reached, all non-spherical structures show higher opacity across the entire spectrum compared to the sphere, though the magnitude of this effect remains wavelength-dependent.

\begin{figure}
    \centering
    \includegraphics[scale=0.48, trim=1.0cm 0mm 0mm 0mm, clip]{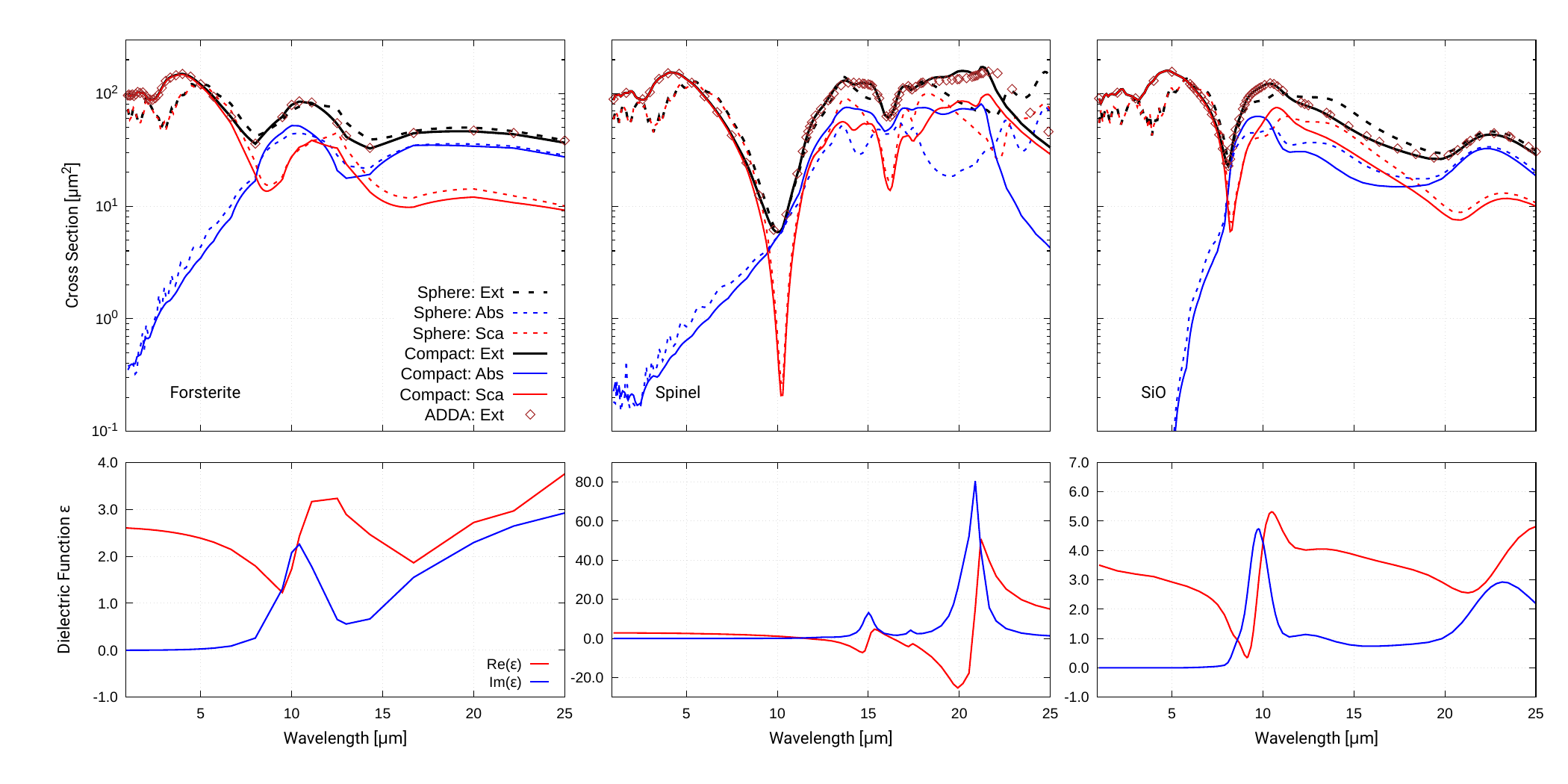}
    \caption{Upper panel: a detailed view on the relative contribution of scattering (red) and absorption (blue) to the extinction cross section (black). From left to right we plot forsterite, spinel and silicon monoxide for a particle of size 3 $\mu$m either spherical (dashed line) or compact aggregate (solid line). In the lower panel we plot the real (red) and imaginary (blue) part of the permittivity for each species. Purple diamonds identify the results obtained running the \texttt{ADDA} code for the same compact ensemble.}
    \label{abs_sca}
\end{figure}

The extinction accounts for the contribution of scattering and absorption; figure \ref{abs_sca} (upper panels) shows (only for the single sphere and the compact cluster) how the main structural differences are due to the scattering cross section below $\sim$7 micron, while beyond this value absorption dominates. The dielectric function of the three materials is shown in the lower panels.

The enhanced scattering cross-section observed for structured particles relative to their equivalent-mass sphere counterparts is primarily driven by their generally larger surface-area-to-volume ratio. Since light scattering is inherently a phenomenon governed by electromagnetic boundary conditions across material interfaces, the higher accessible surface area of an irregular structure naturally enhances it.

Spinel behavior seems to deviate beyond $\lambda\sim 15\,\mu$m: this is driven by its peculiar chemical nature, as it behaves as a highly reflective material in specific infrared windows. In these bands, the material reacts by dramatically increasing its overall polarization. This local electromagnetic boosting enhances both scattering and coupling between nearby particles, producing a scattering driven response that is significantly larger than what is seen in the forsterite or the silicon monoxide case. 

To evaluate the overall discrepancies among structured aggregate models and the standard spherical reference, we analyzed the extinction cross sections as a function of particle equivalent radius. For this analysis, the infrared spectrum was partitioned into three distinct regimes: the short-wave scattering continuum ($1\text{--}7\,\mu\text{m}$), the silicate resonance feature ($7\text{--}12\,\mu\text{m}$), and the long-wave infrared band ($12\text{--}25\,\mu\text{m}$). Within each interval we account for the absolute difference between the single sphere cross sections ($\sigma_{\rm ext,k}^{\rm sp}$) and the structured aggregates ($\sigma_{\rm ext,k}^i$). This parameter represents the mean of the absolute value of the relative difference across the wavelength points $N_{\Delta\lambda}$ within a given band:
\begin{equation}
    d_i(s,\Delta \lambda) = \frac{1}{N_{\Delta\lambda}}\sum_{k=1}^{N_{\Delta \lambda}} \left| \frac{
    \sigma_{\rm ext,k}^i-\sigma_{\rm ext,k}^{\rm sp}}{\sigma_{\rm ext,k}^{\rm sp}} \right|
    \qquad i=\mathrm{com,por,elo}
\end{equation}    
Where $k$ runs through the $N_{\Delta\lambda}$ wavelength points within each of the three intervals defined above. 

\begin{figure}
    \centering
    \includegraphics[scale=0.47, trim=1.5cm 0mm 0mm 0mm, clip]{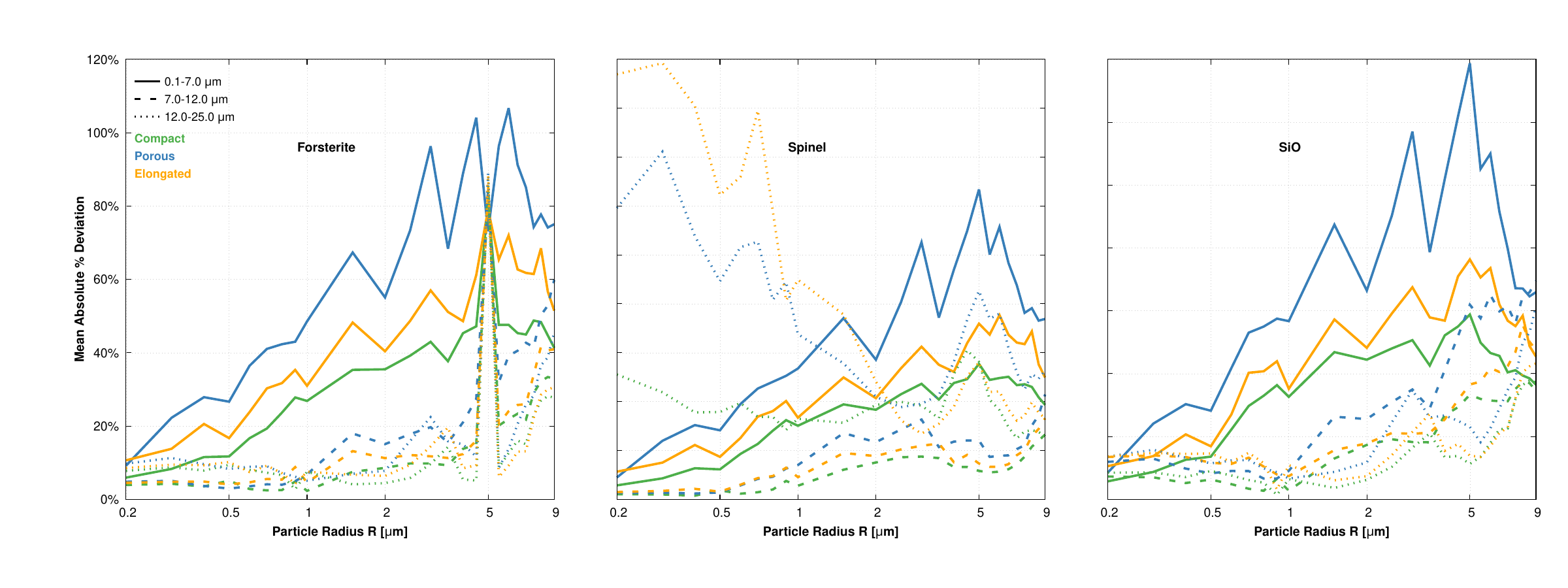}
    \caption{Absolute difference between the various structures (same color coding as fig. \ref{cs_res}) and the single sphere case for forsterite (left), spinel (middle) and silicon monoxide (right). The estimate is done for three wavelength ranges: 1-7 $\mu$m (solid), 7-12 $\mu$m (dashed) and 12-25 $\mu$m (dotted).}
    \label{size_deviation}
\end{figure}

Figure \ref{size_deviation} shows that across all materials, the divergence is most pronounced in the $1\text{--}7\,\mu\text{m}$ band for particles exceeding $5\text{--}6\,\mu\text{m}$ in equivalent radius, where structured models yield extinction values up to a factor of two higher than Mie theory. While these differences are generally moderate at longer wavelengths, spinel represents a notable exception, exhibiting enhanced morphological sensitivity within the $12\text{--}25\,\mu\text{m}$ range. Especially in the near IR regime, these results point towards the necessity of accounting for grain complexity to avoid significant systematic biases in atmospheric modeling.

\section{Condensate chemistry and setup for the fiducial model}\label{P5}
We model the chemistry of a planetary atmosphere using the code $\texttt{cp\_zephyr}$, developed upon the X-ray chemistry code presented in \citet{Locci24}. It has been re--written in modern Fortran, and solves for the equilibrium and kinetic chemistry in a planetary atmosphere in a fully modular structure. Non-equilibrium processes such as vertical mixing and/or photochemistry and ion chemistry can be accounted for as well, together with the possibility of calculating the transmission spectra. 

Furthermore, \texttt{cp\_zephyr} can also solve for the equilibrium condensation following the formalism described in \cite{Kitzmann24} for the code \texttt{FastChem} \citep{Stock18,Stock22}. We do not employ a microphysics model yet to follow the condensate assembly and its emerging physical properties (e.g. mean grain size, dust size distribution etc). 
As a case study for this paper we setup a planet with similar characteristics to HD 189733b with $M_{\rm p}=1.138 M_{\rm J}$, $R_{\rm p} = 1.138 R_{\rm J}$ and stellar radius $R_\star = 0.76 R_{\odot}$. The pressure-temperature profile is taken from \mbox{\cite{Moses11}}, figure 2, left panel, purple line.

In Fig. \ref{val_cpz} we show the validation for the volume mixing ratio of a few key molecules (left panel), and the condensation equilibrium for the solid species described in this paper (middle panel) against \texttt{FastChem}.

In order to obtain a realistic estimate of the amount of condensate that builds up a cloud, we follow the method detailed in \citet{Batalha25} when discussing their \texttt{virga} model, which is based on earlier work by \citet{Ackerman2001}. In their analysis, the abundance of a given species in the gas phase is compared layer by layer to its saturation curve. When condensation occurs, the final amount of material composing the cloud is determined by balancing the upward turbulent mixing of the volatile material against the sedimentation of its condensed phase.

We adopt this framework with a minor modification: we rely on our condensation equilibrium solver to explicitly determine the cloud base at the precise pressure level where solid forsterite first precipitates. Using this same thermodynamic approach, we derive stability curves for forsterite and spinel to compute their corresponding cloud profile. The right panel of Figure \ref{val_cpz} presents our resulting cloud structures, including a comparison that demonstrates excellent agreement with the \texttt{virga} baseline for forsterite. Notably, no silicon monoxide cloud forms, as enstatite (alongside forsterite and diopside) sequesters nearly all available silicon in the upper layers (see also \citeauthor{Visscher10}, \citeyear{Visscher10}).

This approach allows us to retrieve a realistic cloud structure from which we compute the resulting transmission spectrum as detailed in the next section.

The equivalent radius size distribution is assumed to be log-normal with a width of $\sigma=1.5$ while $f_{\rm sed}=1$. We adopt a median grain radius of $r_m = 1\,\mu\text{m}$ as a representative baseline. While the exact median size depends on local atmospheric dynamics, this value allows us to sample the transition region where the breakdown of the spherical approximation is most pronounced. Furthermore, this choice ensures that the bulk of the distribution remains within the $0.1$ to $10\,\mu\text{m}$ range for which our T-matrix cross-sections were explicitly computed, thus avoiding extrapolations into regimes where the morphological impacts are less constrained. Our focus on micron-sized grains is further motivated by the outcome of cloud formation models when investigating silicate clouds in environments similar to the study case we are analyzing here \citep{Powell18}.

This calibration provides a robust framework for our case study, allowing us to quantify the spectral impact of adopting complex, structured grain morphologies as an alternative to the traditional spherical approximation in planetary transmission spectroscopy.

\section{Transmission spectrum}\label{P6}

The transmission spectrum is observed when the planet is transiting in front of the star and is mathematically defined as the transit depth at each wavelength:
\begin{equation}
    \delta_\lambda = \frac{R_{\rm pl}^2+A(\lambda)}{R_*^2}
\end{equation}
where the equivalent depth $A(\lambda)$ accounts for the extinction within the planetary atmosphere:
\begin{equation}
    A(\lambda) = 2\int_{0}^{z_{\rm max}}(R_{\rm pl}+z)\left(1-e^{-\tau(\lambda,z)}\right)dz
\end{equation}
where the integration is carried out along the atmosphere's altimetric profile $z$. All the extinction properties of the medium are evaluated through the function $\tau(\lambda,z)$ which varies according to a given layer and its local chemistry, being naturally impacted, e.g. by the presence of condensates. The radiative transfer is evaluated in a 1D grid as detailed in \citet{Locci24}.

The transmission spectrum is computed across the 1–10 $\mu$m range, covering the operational window of JWST and Ariel. We incorporate gas-phase molecular data from the Exomol database \citep{Tennyson16} at a spectral resolution of $R \sim 15,000$, although the resulting spectrum is degraded to a nominal R$\sim$300, similar to what the Ariel instrument is expected to achieve \citep{Tinetti18}.

Here, we aim at isolating the quantitative contribution of each grain aggregation geometry. Although the impact on the total spectrum is strongly model-dependent (e.g. temperature-pressure profile, scale height $H$, $k_{zz}$, $f_{\rm sed}$ etc.), the relative contribution of each morphology provides a clear comparative baseline.

To isolate the specific impact of grain morphology on the observable spectrum, we include only the extinction cross-sections for forsterite, spinel, and SiO calculated in this work. By intentionally excluding other potential condensate opacities, we ensure that any observed spectral deviations can be attributed directly to the transition from spherical to complex grain geometries.

\begin{figure}
    \centering
    \includegraphics[scale=0.65]{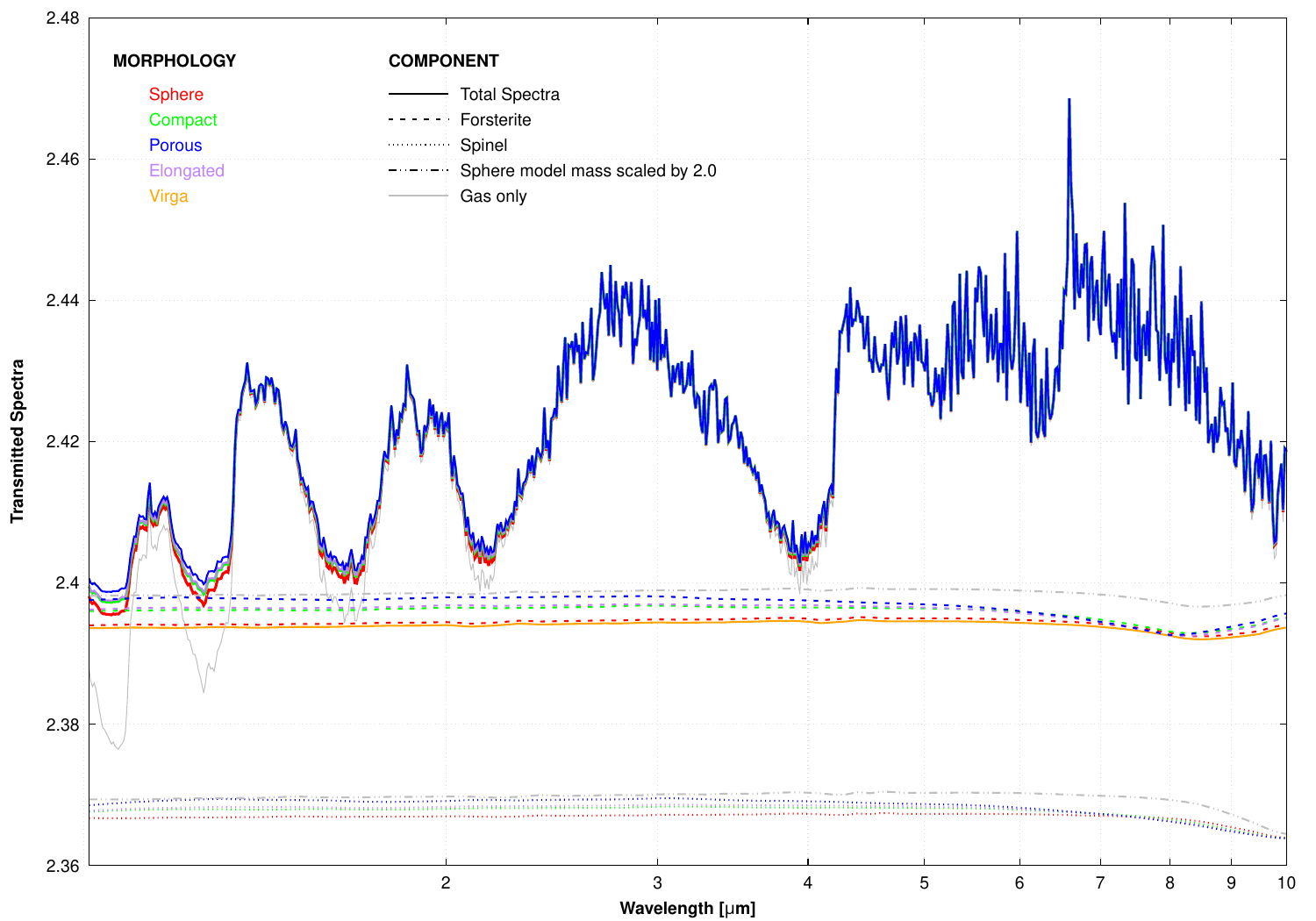}
    \caption{Transmission spectrum for the fiducial planet between 1 and 10 microns. The color coding for the different structures is shown in the top left. In orange we show the contribution to the transmission spectrum from forsterite by using the output distribution of the condensate obtained with \texttt{virga}. The contribution to the transmission spectrum is shown for forsterite (dashed) and spinel (dotted) together with a model with scaled up dust mass (factor 2.0) for the single sphere case (dot-dashed, grey). Solid grey indicates what the transmission spectrum would be ignoring the presence of condensates.}
    \label{trspec}
\end{figure}

The resulting transmission spectrum is shown in Figure \ref{trspec}. The total spectrum (colored solid lines) is primarily dominated by gas-phase absorption, driven predominantly by H$_2$O with additional contributions from CH$_4$, CO$_2$, and CO. A cloud-free, gas-only model is shown for reference (solid grey line). Against this gas-phase background, the condensates introduce a grey opacity continuum, the exact level of which depends on our specific assumptions regarding the nature of the cloud particles.

Focusing on what the differences are between the condensates' grey absorption among different aggregation modes, we can observe that the impact on the contribution of each condensate is significant. As illustrated in Fig. \ref{trspec}, the transition from the spherical baseline to complex morphologies is characterized by a progressive increase in continuum opacity. This effect follows a clear hierarchy of structural complexity, with extinction efficiency rising as we move from single spheres to compact and elongated structures, reaching its maximum in the porous ensembles (this is color coded as indicated in the morphology legend in the plot). Moreover, and more importantly, this effect seems to apply independently of the material composing the solid grains.

As a qualitative diagnostic, we also include a model utilizing spherical grains with their opacity scaled by a factor of 2.0. As shown in Fig. \ref{trspec}, the resulting transmission features for both forsterite and spinel (dot-dashed) at wavelengths below $5$ $\mu\text{m}$ are remarkably similar to those of the porous ensemble. This overlap highlights a significant degeneracy: without accounting for grain morphology, the opacity can be mistakenly attributed to a higher condensate mass, essentially almost doubling the inferred dust abundance to match the observed spectral depth.

This difference can be important and should be taken into account by forward or retrieval models.

\section{Discussion and conclusions}\label{P7}
Recently, several studies have investigated the impact of non-spherical geometries on atmospheric scattering, exploring morphologies such as fractal aggregates \citep{Moran25}, crystalline cuboids for KCl \citep{Hamill25}, but also fluffy aggregates originating from photochemistry (hazes) with different porosity in HD 189733b \citep{Vahidinia24} or aggregates of spherical monomers building KCl solid condensates in the atmosphere of GJ 1214b \citep{Ohno20}.

Similarly, in this work we investigate various modes of grain aggregation across a wide range of equivalent radii.

For small particles our findings are in qualitative agreement with the results of \citet{Tazaki18}, who performed a comparative analysis between accelerated scattering algorithms and rigorous T-matrix calculations. Specifically, in their investigation, they observed that Mie theory tends to overestimate absorption at near-infrared wavelengths when compared to T-matrix results for structured aggregates with dimension up to 1 micron. Furthermore, the scattering cross-sections reported in their study exhibit a behavior remarkably similar to our own models, reinforcing the conclusion that spherical approximations fail to capture the complex interference effects inherent in non-spherical mineral condensates.

In the same size regime we also find our results in agreement with \citet{Silsbee16} as they find a lower extinction cross section (and a resulting lower radiation pressure) for porous grains when compared to the single sphere case.

We furthermore explored grain radii exceeding $1\,\mu\text{m}$, observing a widening divergence between Mie theory and our T-matrix results. In this regime, the spherical approximation consistently yields a systematic underestimation of the total extinction cross-section. This discrepancy suggests that multiple-scattering processes become increasingly relevant once the aggregate size becomes comparable to the incident wavelength, because electromagnetic coupling among constituent monomers is most pronounced. A larger surface-to-volume ratio in the aggregate compared with the equivalent sphere reflects a more complex morphology with a greater number of internal interfaces, promoting stronger electromagnetic coupling among the constituent monomers. 

In atmospheric data analysis, cloud modeling is treated with varying levels of complexity depending on the required computational efficiency. In retrieval modeling, the need for rapid parameter space exploration necessitates a parameterized approach to cloud structure and opacity, aiming to identify statistically robust values for the atmospheric state \citep[e.g.,][]{Dyrek23,Beatty24}. Conversely, forward modeling employs a diverse array of microphysical and chemical models to simulate the actual formation of clouds and their resulting impact on transmission spectra, seeking a physical setup that best matches the observed data, see \citet{Grant23} and \citet{Rustamkulov23}. Despite these methodological differences, both approaches typically rely on the fundamental assumption that condensates are spherical, utilizing Mie theory to calculate their extinction properties. 

In this work, we demonstrate that the opacity of structured particles can be a factor of 2.0 to 3.0 higher than that of equivalent-mass spheres. In our fiducial planetary models, this enhanced extinction translates into a significantly higher contribution to the transmission spectrum. Consequently, we estimate that models relying on the spherical approximation could systematically overestimate the required condensate mass by up to a factor of two to achieve the same level of observed absorption.

This degeneracy has profound implications for the interpretation of exoplanetary spectra in the high-precision era of \textit{JWST} and the upcoming \textit{Ariel} mission. Overestimating the required aerosol mass to fit a muted transmission spectrum impacts for example on the derived elemental abundances, as a larger fraction of refractory elements (such as Mg, Si, and O) would be erroneously assumed to be locked away in the solid phase. Consequently, transitioning away from the spherical approximation is not merely a microphysical refinement, but a necessary step to accurately constrain atmospheric metallicities, C/O ratios, and ultimately, planetary formation histories.

The application of the T-matrix algorithm provides a robust methodology for calculating realistic extinction cross-sections for the mineral condensates increasingly associated with grey absorption in planetary transmission spectra. 

In future work, we intend to expand our database by incorporating additional species, such as diopside (CaMgSi$_2$O$_6$), SiO$_2$, and enstatite (MgSiO$_3$), while extending the spectral range of our calculations towards lower wavelengths and possibly the size of the aggregate itself. Furthermore, we will investigate more diverse dust morphologies, including heterogeneous aggregates composed of varying monomer sizes. Finally, we plan to couple our exact optical calculations with a comprehensive microphysical model to explicitly track the nucleation and coagulation of condensates. This dynamical approach will allow us to self-consistently derive the emergent size distributions and physical structures of the particles, significantly reducing our reliance on parameterized assumptions and further refining our understanding of how complex grains influence atmospheric observables.

The detailed structure of each aggregation mode and the calculated cross sections are publicly available through Zenodo at the following links: forsterite (https://doi.org/10.5281/zenodo.19231040), spinel (https://doi.org/10.5281/zenodo.19330375) and silicon monoxide (https://doi.org/10.5281/zenodo.19330602). This repository will be continuously expanded as new species are computed, providing the exoplanet community with a robust, open-access library of exact extinction cross-sections for complex atmospheric aggregates.

\section*{acknowledgments}
Supported by the Italian Ministry of University and Research (MUR), through PRIN Exo-CASH (PRIN 2022 project no. 2022J7ZFRA). The NP\_TMcode software is the parallel implementation of the T-matrix formalism, developed and maintained at INAF - Astronomical Observatory
of Cagliari as a contribution to the activities of ICSC Spoke 3 “Astrophysics and Cosmos Observations”, on the basis of the codes originally written by F. Borghese, P. Denti and R. Saija. It is registered in the Astrophysics Source Code
Library at https://ascl.net/2510.003. 

The models for this paper were calculated on the computing Cluster at the Osservatorio Astronomico di Cagliari, funded by the National Operational Program (PON) of the Italian Ministry of University and Research “Research and Innovation 2014-2020”, Notice D.D. 424 of 02/28/2018 for granting funds aimed at enhancing research infrastructures, implementing Action II.1 – Project Proposal PIR01\_00010.

A. P. and G. A. acknowledge support from INAF project 1.05.23.04.02. – MINI-GRANTS (2023) di RSN2 “Modelling of exoplanetary atmospheres in different physical conditions”.

M. A. Iat\`i and F. Saija acknowledge Project ”Space It Up” funded by ASI and MUR– Contract n. 2024-5-E.0 - CUP n. I53D24000060005.

\appendix

\setcounter{figure}{0}
\renewcommand{\thefigure}{A\arabic{figure}}

\setcounter{table}{0}
\renewcommand{\thetable}{A\arabic{table}}
\section{Validation of the chemical model}
\begin{figure}[h!]
    \centering
\includegraphics[scale=0.73]{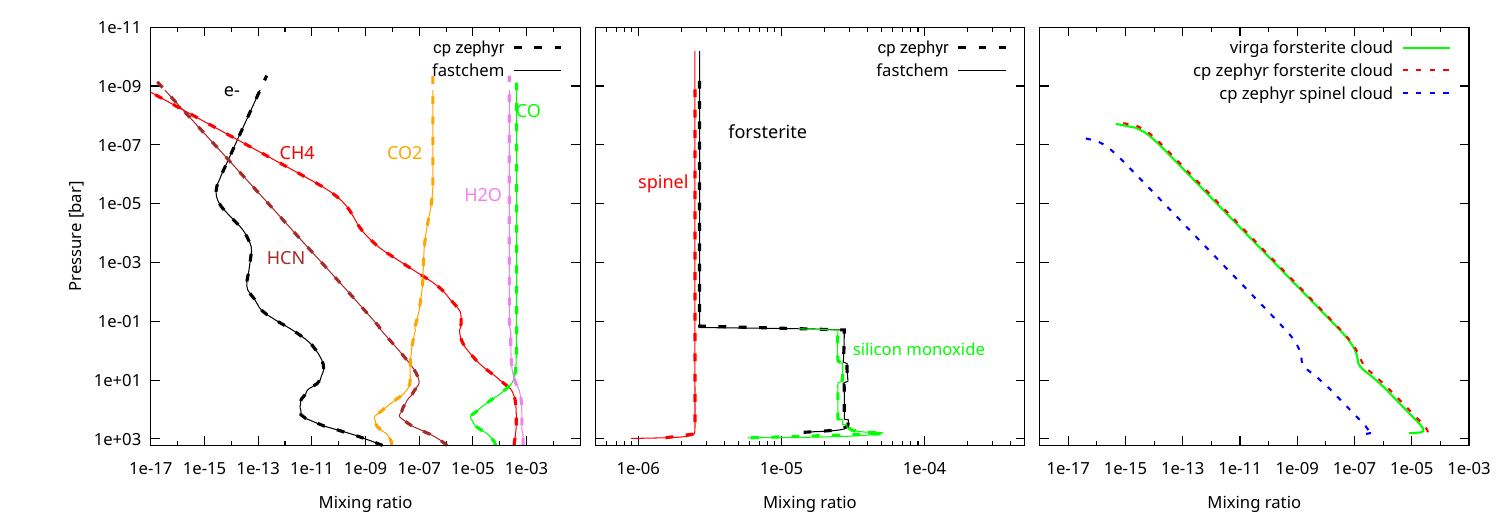}
\caption{Left plot: validation for the gas phase in the equilibrium chemistry case for our code \texttt{cp\_zephyr} (dashed) and \texttt{FastChem} (solid). Middle plot: validation when condensation is switched on for the condensates studied here. Right plot: the cloud shapes for forsterite and spinel (dashed) vs the one obtained with Virga for a similar atmosphere (solid).}
    \label{val_cpz}
\end{figure}

We validate our chemical model against \texttt{FastChem} in Figure \ref{val_cpz}. The left panel shows the gas-only equilibrium for a subset of species, while the middle panel demonstrates the benchmark when condensation is enabled for the key condensates in this paper. An extensive validation (not shown here) has been performed for different pressure-temperature profiles and metallicities.

\section{Specifics on grain configurations} 
We provide here a table summarizing the values of the grains sizes in each aggregation mode, also specifying the number of spheres used and the size of the monomer chosen as a building block.
\begin{table}[H]
\centering
\begin{tabular}{|c|c|c|c|c|c|c|}
 \hline
   $r_{\mathrm eq}$ [$\mu$m]  & $N_{\mathrm{sph}}$ & $r_{\mathrm enc}$ compact [$\mu$m] & $r_{\mathrm enc}$  porous [$\mu$m] &$r_{\mathrm enc}$ elongated [$\mu$m] & Monomer size [$\mu$m] & Wall-clock time [s]\\ 
 \hline  
 0.1 & 1 & 0.1 & 0.1 & 0.1 & 0.10 & 10\\
0.2 & 8 & 0.32 & 0.49 & 0.50 & 0.10    &  13     \\
 0.3 & 27 & 0.44 & 0.89 & 0.71 & 0.10  &  21     \\
 0.4 & 64 & 0.58 & 1.15 & 0.89 & 0.10  &  44     \\
 0.5 & 16 & 0.78 & 1.08  & 1.10 & 0.20 &  30     \\
 0.6 & 27 & 0.89 & 1.77 & 1.40 & 0.20  &  33     \\
 0.7 & 43 & 1.01 & 2.28 & 1.82 & 0.20  &  47     \\
 0.8 & 64 & 1.15 & 1.82 & 1.77 & 0.20  &  68     \\
 0.9 & 11 & 1.27 & 2.70 & 2.00 & 0.40  &  40     \\
 1.0 & 16 & 1.57 & 2.17 & 2.21 & 0.40  &  48     \\
 1.5 & 53 & 2.17 & 5.23 & 4.00 & 0.40  &  91     \\
 2.0 & 16 & 3.13 & 4.19 & 4.41 & 0.80  &  195    \\
 2.5 & 31 & 3.73 & 6.41 & 5.60 & 0.80  &  706    \\
 3.0 & 53 & 4.35 & 8.78 & 7.20 & 0.80  &  2336   \\
 3.5 & 10 & 5.08 & 8.48 & 7.90 & 1.60  &  3100   \\
 4.0 & 16 & 6.27 & 10.6 & 8.82 & 1.60  &  4690   \\
 4.5 & 22 & 6.79 & 16.1 & 11.3 & 1.60  &  6788   \\
 5.0 & 31 & 7.47 & 15.2 & 11.5 & 1.60  &  11361  \\
 5.5 & 41 & 8.03 & 14.3 & 12.7 & 1.60  &  8135   \\
 6.0 & 13 & 9.18 & 17.0 & 13.8 & 2.55  &  8930   \\
 6.5 & 12 & 9.67 & 13.6 & 15.0 & 2.84  &  8109   \\
 7.0 & 10 & 10.16 & 14.7 & 15.8 & 3.20 &  9085   \\
 7.5 & 10 & 11.09 & 14.8 & 17.3 & 3.49 &  10521  \\
 8.0 & 10 & 11.89 & 18.7 & 18.5 & 3.74 &  13658  \\
 8.5 & 10 & 12.59 & 18.5 & 19.7 & 3.96 &  14655  \\
 9.0 & 10 & 13.22 & 19.4 & 20.6 & 4.16 &  15440  \\
 9.5 & 10 & 13.98 & 22.1 & 21.8 & 4.40 &  16686  \\
 10.0 & 10 & 14.82 & 21.1 & 23.1 & 4.66 & 23768  \\
 \hline
\end{tabular}
\caption{Setup for the structures studied in this paper. From left to right: radius of the spherical particles of a given mass, number of spheres used in the structured models, radius encircling the whole structure for the three aggregation modes, size of the monomer. In each structured case the mass of the grain is the same as the one contained in the spherical one. The last column shows the wall-clock time for the silicon monoxide calculations in compact mode for 241 wavelength points. The calculation times for other aggregation modes are on average less than a factor two longer. The RAM usage for these models goes from a fraction of GiB to about 280 GiB for the largest particle. The RAM needed depends mostly on the particle's equivalent radius, the number of monomers and the number of lambda points being computed in parallel.}
\label{size_table}
\end{table}

\bibliography{particles}{}

\begin{thebibliography}{}
\expandafter\ifx\csname natexlab\endcsname\relax\def\natexlab#1{#1}\fi
\providecommand{\url}[1]{\href{#1}{#1}}
\providecommand{\dodoi}[1]{doi:~\href{http://doi.org/#1}{\nolinkurl{#1}}}
\providecommand{\doeprint}[1]{\href{http://ascl.net/#1}{\nolinkurl{http://ascl.net/#1}}}
\providecommand{\doarXiv}[1]{\href{https://arxiv.org/abs/#1}{\nolinkurl{https://arxiv.org/abs/#1}}}

\bibitem[{Ackerman \& Marley(2001)}]{Ackerman2001}
Ackerman, A.~S., \& Marley, M.~S. 2001, The Astrophysical Journal, 556, 872, \dodoi{10.1086/321540}

\bibitem[{Adams {et~al.}(2019)Adams, Gao, Pater, \& Morley}]{Adams19}
Adams, D., Gao, P., Pater, I.~d., \& Morley, C.~V. 2019, The Astrophysical Journal, 874, 61, \dodoi{10.3847/1538-4357/ab074c}

\bibitem[{{Baburaj} {et~al.}(2026){Baburaj}, {Ruffio}, {Perrin}, {Xuan}, {Balmer}, {Chachan}, {Konopacky}, {Barman}, {M{\^a}lin}, {Hoch}, {Rickman}, {Ward-Duong}, {Pueyo}, {Girard}, {Rebollido}, {Bidot}, {Chen}, {Worthen}, {Lu}, {Kammerer}, {van der Marel}, {Lewis}, {Valenti}, {Seager}, {Stark}, {Soummer}, {Anderson}, {Lajoie}, {Clampin}, \& {Mountain}}]{Baburaj26}
{Baburaj}, A., {Ruffio}, J.-B., {Perrin}, M., {et~al.} 2026, arXiv e-prints, arXiv:2606.19228.
\newblock \doarXiv{2606.19228}

\bibitem[{{Barman} {et~al.}(2015){Barman}, {Konopacky}, {Macintosh}, \& {Marois}}]{Barman15}
{Barman}, T.~S., {Konopacky}, Q.~M., {Macintosh}, B., \& {Marois}, C. 2015, \apj, 804, 61, \dodoi{10.1088/0004-637X/804/1/61}

\bibitem[{Batalha {et~al.}(2025)Batalha, Rooney, Visscher, Moran, Marley, Sengupta, Kiefer, Lodge, Mang, Morley, Mukherjee, Fortney, Gao, Lewis, Mayorga, Pearce, \& Wakeford}]{Batalha25}
Batalha, N.~E., Rooney, C.~M., Visscher, C., {et~al.} 2025, Condensation Clouds in Substellar Atmospheres with Virga.
\newblock \doarXiv{2508.15102}

\bibitem[{Beatty {et~al.}(2024)Beatty, Welbanks, Schlawin, Bell, Line, Murphy, Edelman, Greene, Fortney, Henry, Mukherjee, Ohno, Parmentier, Rauscher, Wiser, \& Triantafillides}]{Beatty24}
Beatty, T.~G., Welbanks, L., Schlawin, E., {et~al.} 2024, The Astrophysical Journal Letters, 970, L10, \dodoi{10.3847/2041-8213/ad55e9}

\bibitem[{Borghese {et~al.}(2007)Borghese, Denti, \& Saija}]{Borghese07}
Borghese, F., Denti, P., \& Saija, R. 2007, Scattering from Model Nonspherical Particles: Theory and Applications to Environmental Physics (Berlin, Heidelberg: Springer)

\bibitem[{Borghese {et~al.}(1980)Borghese, Denti, Toscano, \& Sindoni}]{Borghese80}
Borghese, F., Denti, P., Toscano, G., \& Sindoni, O.~I. 1980, Journal of Mathematical Physics, 21, 2754, \dodoi{10.1063/1.524394}

\bibitem[{Carone {et~al.}(2026)Carone, Helling, Gernjak, Leitner, Janz, Iro, Kiefer, Min, Samra, \& Woitke}]{Carone26}
Carone, L., Helling, C., Gernjak, S., {et~al.} 2026, \aap, 708, A185, \dodoi{10.1051/0004-6361/202556695}

\bibitem[{{Changeat} {et~al.}(2024){Changeat}, {Skinner}, {Cho}, {N{\"a}ttil{\"a}}, {Waldmann}, {Al-Refaie}, {Dyrek}, {Edwards}, {Mikal-Evans}, {Joshua}, {Morello}, {Skaf}, {Tsiaras}, {Venot}, \& {Yip}}]{Changeat24}
{Changeat}, Q., {Skinner}, J.~W., {Cho}, J.~Y.~K., {et~al.} 2024, \apjs, 270, 34, \dodoi{10.3847/1538-4365/ad1191}

\bibitem[{{de Kok} {et~al.}(2013){de Kok}, {Brogi}, {Snellen}, {Birkby}, {Albrecht}, \& {de Mooij}}]{deKok13}
{de Kok}, R.~J., {Brogi}, M., {Snellen}, I.~A.~G., {et~al.} 2013, \aap, 554, A82, \dodoi{10.1051/0004-6361/201321381}

\bibitem[{Draine \& Flatau(1994)}]{Draine94}
Draine, B.~T., \& Flatau, P.~J. 1994, Journal of the Optical Society of America A, 11, 1491, \dodoi{10.1364/JOSAA.11.001491}

\bibitem[{Dyrek {et~al.}(2023)Dyrek, Min, Decin, Bouwman, Crouzet, Mollière, Lagage, Konings, Tremblin, Güdel, Pye, Waters, Henning, Vandenbussche, Ardevol~Martinez, Argyriou, Ducrot, Heinke, van Looveren, Absil, Barrado, Baudoz, Boccaletti, Cossou, Coulais, Edwards, Gastaud, Glasse, Glauser, Greene, Kendrew, Krause, Lahuis, Mueller, Olofsson, Patapis, Rouan, Royer, Scheithauer, Waldmann, Whiteford, Colina, van Dishoeck, Östlin, Ray, \& Wright}]{Dyrek23}
Dyrek, A., Min, M., Decin, L., {et~al.} 2023, Nature, 625, 51–54, \dodoi{10.1038/s41586-023-06849-0}

\bibitem[{{Evans} {et~al.}(2016){Evans}, {Sing}, {Wakeford}, {Nikolov}, {Ballester}, {Drummond}, {Kataria}, {Gibson}, {Amundsen}, \& {Spake}}]{Evans16}
{Evans}, T.~M., {Sing}, D.~K., {Wakeford}, H.~R., {et~al.} 2016, \apjl, 822, L4, \dodoi{10.3847/2041-8205/822/1/L4}

\bibitem[{{Fabian} {et~al.}(2001){Fabian}, {Posch}, {Mutschke}, {Kerschbaum}, \& {Dorschner}}]{Fabian01}
{Fabian}, D., {Posch}, T., {Mutschke}, H., {Kerschbaum}, F., \& {Dorschner}, J. 2001, \aap, 373, 1125, \dodoi{10.1051/0004-6361:20010657}

\bibitem[{{Gao} {et~al.}(2021){Gao}, {Wakeford}, {Moran}, \& {Parmentier}}]{Gao21}
{Gao}, P., {Wakeford}, H.~R., {Moran}, S.~E., \& {Parmentier}, V. 2021, Journal of Geophysical Research (Planets), 126, e06655, \dodoi{10.1029/2020JE006655}

\bibitem[{{Grant} {et~al.}(2023){Grant}, {Lewis}, {Wakeford}, {Batalha}, {Glidden}, {Goyal}, {Mullens}, {MacDonald}, {May}, {Seager}, {Stevenson}, {Valenti}, {Visscher}, {Alderson}, {Allen}, {Ca{\~n}as}, {Col{\'o}n}, {Clampin}, {Espinoza}, {Gressier}, {Huang}, {Lin}, {Long}, {Louie}, {Pe{\~n}a-Guerrero}, {Ranjan}, {Sotzen}, {Valentine}, {Anderson}, {Balmer}, {Bellini}, {Hoch}, {Kammerer}, {Libralato}, {Mountain}, {Perrin}, {Pueyo}, {Rickman}, {Rebollido}, {Sohn}, {van der Marel}, \& {Watkins}}]{Grant23}
{Grant}, D., {Lewis}, N.~K., {Wakeford}, H.~R., {et~al.} 2023, \apjl, 956, L32, \dodoi{10.3847/2041-8213/acfc3b10.3847/2041-8213/acfdab}

\bibitem[{Hamill {et~al.}(2025)Hamill, Johnson, Lodge, Gao, Nag, Batalha, Christie, \& Wakeford}]{Hamill25}
Hamill, C.~D., Johnson, A.~V., Lodge, M., {et~al.} 2025, The Astrophysical Journal, 987, 176, \dodoi{10.3847/1538-4357/add7cf}

\bibitem[{Hass \& Salzberg(1954)}]{Hass54}
Hass, G., \& Salzberg, C.~D. 1954, J. Opt. Soc. Am., 44, 181, \dodoi{10.1364/JOSA.44.000181}

\bibitem[{Herguedas \& Carretero(2023)}]{Herguedas23}
Herguedas, N., \& Carretero, E. 2023, Nanomaterials, 13, \dodoi{10.3390/nano13202749}

\bibitem[{{J{\"a}ger} {et~al.}(2003){J{\"a}ger}, {Dorschner}, {Mutschke}, {Posch}, \& {Henning}}]{Jager03}
{J{\"a}ger}, C., {Dorschner}, J., {Mutschke}, H., {Posch}, T., \& {Henning}, T. 2003, \aap, 408, 193, \dodoi{10.1051/0004-6361:20030916}

\bibitem[{{Kempton} {et~al.}(2023){Kempton}, {Zhang}, {Bean}, {Steinrueck}, {Piette}, {Parmentier}, {Malsky}, {Roman}, {Rauscher}, {Gao}, {Bell}, {Xue}, {Taylor}, {Savel}, {Arnold}, {Nixon}, {Stevenson}, {Mansfield}, {Kendrew}, {Zieba}, {Ducrot}, {Dyrek}, {Lagage}, {Stassun}, {Henry}, {Barman}, {Lupu}, {Malik}, {Kataria}, {Ih}, {Fu}, {Welbanks}, \& {McGill}}]{Kempton23}
{Kempton}, E. M.-R., {Zhang}, M., {Bean}, J.~L., {et~al.} 2023, \nat, 620, 67, \dodoi{10.1038/s41586-023-06159-5}

\bibitem[{{Kitzmann} {et~al.}(2024){Kitzmann}, {Stock}, \& {Patzer}}]{Kitzmann24}
{Kitzmann}, D., {Stock}, J.~W., \& {Patzer}, A. B.~C. 2024, \mnras, 527, 7263, \dodoi{10.1093/mnras/stad3515}

\bibitem[{{Knutson} {et~al.}(2014){Knutson}, {Benneke}, {Deming}, \& {Homeier}}]{Knutson14}
{Knutson}, H.~A., {Benneke}, B., {Deming}, D., \& {Homeier}, D. 2014, \nat, 505, 66, \dodoi{10.1038/nature12887}

\bibitem[{{Kreidberg} {et~al.}(2014){Kreidberg}, {Bean}, {D{\'e}sert}, {Benneke}, {Deming}, {Stevenson}, {Seager}, {Berta-Thompson}, {Seifahrt}, \& {Homeier}}]{Kreidberg14}
{Kreidberg}, L., {Bean}, J.~L., {D{\'e}sert}, J.-M., {et~al.} 2014, \nat, 505, 69, \dodoi{10.1038/nature12888}

\bibitem[{{La Mura} {et~al.}(2026){La Mura}, {Mulas}, {Aresu}, {Iat{\`\i}}, {Cecchi-Pestellini}, {Rezaei}, \& {Saija}}]{LaMura26}
{La Mura}, G., {Mulas}, G., {Aresu}, G., {et~al.} 2026, Astronomy and Computing, 55, 101077, \dodoi{10.1016/j.ascom.2026.101077}

\bibitem[{{Locci} {et~al.}(2024){Locci}, {Aresu}, {Petralia}, {Micela}, {Maggio}, \& {Cecchi-Pestellini}}]{Locci24}
{Locci}, D., {Aresu}, G., {Petralia}, A., {et~al.} 2024, \psj, 5, 58, \dodoi{10.3847/PSJ/ad24e4}

\bibitem[{{Lodge} {et~al.}(2024){Lodge}, {Wakeford}, \& {Leinhardt}}]{Lodge24}
{Lodge}, M.~G., {Wakeford}, H.~R., \& {Leinhardt}, Z.~M. 2024, \mnras, 527, 11113, \dodoi{10.1093/mnras/stad3743}

\bibitem[{{MacDonald}(2023)}]{MacDon23}
{MacDonald}, R.~J. 2023, The Journal of Open Source Software, 8, 4873, \dodoi{10.21105/joss.04873}

\bibitem[{Marley {et~al.}(2013)Marley, Ackerman, Cuzzi, \& Kitzmann}]{Marley13}
Marley, M.~S., Ackerman, A.~S., Cuzzi, J.~N., \& Kitzmann, D. 2013, Clouds and hazes in exoplanet atmospheres, Space Science Series (Phoenix: University of Arizona Press), 367--391, \dodoi{10.2458/azu_uapress_9780816530595-ch015}

\bibitem[{{Mehta, Nishil} {et~al.}(2026){Mehta, Nishil}, {Parmentier, Vivien}, {Tan, Xianyu}, {Lee, Elspeth K. H.}, {Guillot, Tristan}, {Wiser, Lindsey S.}, {Bell, Taylor J.}, {Schlawin, Everett}, {Arnold, Kenneth}, {Mukherjee, Sagnick}, {Greene, Thomas P.}, {Beatty, Thomas G.}, {Welbanks, Luis}, {Line, Michael R.}, {Murphy, Matthew M.}, {Fortney, Jonathan J.}, \& {Ohno, Kazumasa}}]{Mehta26}
{Mehta, Nishil}, {Parmentier, Vivien}, {Tan, Xianyu}, {et~al.} 2026, \aap, 707, A394, \dodoi{10.1051/0004-6361/202557411}

\bibitem[{Mollière {et~al.}(2019)Mollière, Wardenier, van Boekel, Henning, Molaverdikhani, \& Snellen}]{Molliere19}
Mollière, P., Wardenier, J.~P., van Boekel, R., {et~al.} 2019, \aap, 627, A67, \dodoi{10.1051/0004-6361/201935470}

\bibitem[{Moran {et~al.}(2025)Moran, Lodge, Batalha, Ohno, Vahidinia, Marley, Wakeford, \& Leinhardt}]{Moran25}
Moran, S.~E., Lodge, M.~G., Batalha, N.~E., {et~al.} 2025, The Astrophysical Journal, 994, 116, \dodoi{10.3847/1538-4357/ae0583}

\bibitem[{Moses {et~al.}(2011)Moses, Visscher, Fortney, Showman, Lewis, Griffith, Klippenstein, Shabram, Friedson, Marley, \& Freedman}]{Moses11}
Moses, J.~I., Visscher, C., Fortney, J.~J., {et~al.} 2011, The Astrophysical Journal, 737, 15, \dodoi{10.1088/0004-637x/737/1/15}

\bibitem[{{Mukherjee} {et~al.}(2026){Mukherjee}, {Sing}, {Fu}, {Stevenson}, {Schmidt}, {Baskett}, {Mak}, {McCreery}, {Allen}, {Bennett}, {Christie}, {Gasc{\'o}n}, {Goyal}, {H{\'e}brard}, {Lothringer}, {L{\'o}pez-Morales}, {Lustig-Yaeger}, {May}, {Mayorga}, {Mayne}, {Ramos Rosado}, {Reggiani}, {Rustamkulov}, {Schlaufman}, {Sotzen}, {Thorngren}, {Wang}, \& {Zamyatina}}]{Mukherjee26}
{Mukherjee}, S., {Sing}, D.~K., {Fu}, G., {et~al.} 2026, Science, 392, 858, \dodoi{10.1126/science.adx5903}

\bibitem[{Mullens {et~al.}(2024)Mullens, Lewis, \& MacDonald}]{Mullens24}
Mullens, E., Lewis, N.~K., \& MacDonald, R.~J. 2024, Implementation of Aerosol Mie Scattering in POSEIDON with Application to the hot Jupiter HD 189733 b's Transmission, Emission, and Reflected Light Spectrum.
\newblock \doarXiv{2410.19253}

\bibitem[{Ohno {et~al.}(2020)Ohno, Okuzumi, \& Tazaki}]{Ohno20}
Ohno, K., Okuzumi, S., \& Tazaki, R. 2020, The Astrophysical Journal, 891, 131, \dodoi{10.3847/1538-4357/ab44bd}

\bibitem[{Petrova {et~al.}(2000)Petrova, Jockers, \& Kiselev}]{Petrova2000}
Petrova, E.~V., Jockers, K., \& Kiselev, N.~N. 2000, Icarus, 148, 526, \dodoi{https://doi.org/10.1006/icar.2000.6504}

\bibitem[{Philipp(1971)}]{Philipp71}
Philipp, H. 1971, Journal of Physics and Chemistry of Solids, 32, 1935, \dodoi{https://doi.org/10.1016/S0022-3697(71)80159-2}

\bibitem[{Powell {et~al.}(2018)Powell, Zhang, Gao, \& Parmentier}]{Powell18}
Powell, D., Zhang, X., Gao, P., \& Parmentier, V. 2018, The Astrophysical Journal, 860, 18, \dodoi{10.3847/1538-4357/aac215}

\bibitem[{Rustamkulov {et~al.}(2023)Rustamkulov, Sing, Mukherjee, May, Kirk, Schlawin, Line, Piaulet, Carter, Batalha, Goyal, López-Morales, Lothringer, MacDonald, Moran, Stevenson, Wakeford, Espinoza, Bean, Batalha, Benneke, Berta-Thompson, Crossfield, Gao, Kreidberg, Powell, Cubillos, Gibson, Leconte, Molaverdikhani, Nikolov, Parmentier, Roy, Taylor, Turner, Wheatley, Aggarwal, Ahrer, Alam, Alderson, Allen, Banerjee, Barat, Barrado, Barstow, Bell, Blecic, Brande, Casewell, Changeat, Chubb, Crouzet, Daylan, Decin, Désert, Mikal-Evans, Feinstein, Flagg, Fortney, Harrington, Heng, Hong, Hu, Iro, Kataria, Kempton, Krick, Lendl, Lillo-Box, Louca, Lustig-Yaeger, Mancini, Mansfield, Mayne, Miguel, Morello, Ohno, Palle, Petit dit de~la Roche, Rackham, Radica, Ramos-Rosado, Redfield, Rogers, Shkolnik, Southworth, Teske, Tremblin, Tucker, Venot, Waalkes, Welbanks, Zhang, \& Zieba}]{Rustamkulov23}
Rustamkulov, Z., Sing, D.~K., Mukherjee, S., {et~al.} 2023, Nature, 614, 659–663, \dodoi{10.1038/s41586-022-05677-y}

\bibitem[{{Saija} {et~al.}(2001){Saija}, {Iat{\`\i}}, {Borghese}, {Denti}, {Aiello}, \& {Cecchi-Pestellini}}]{Saija01}
{Saija}, R., {Iat{\`\i}}, M.~A., {Borghese}, F., {et~al.} 2001, The Astrophysical Journal, 559, 993, \dodoi{10.1086/322350}

\bibitem[{Shen {et~al.}(2008)Shen, Draine, \& Johnson}]{Shen2008}
Shen, Y., Draine, B.~T., \& Johnson, E.~T. 2008, The Astrophysical Journal, 689, 260, \dodoi{10.1086/592765}

\bibitem[{{Silsbee} \& {Draine}(2016)}]{Silsbee16}
{Silsbee}, K., \& {Draine}, B.~T. 2016, \apj, 818, 133, \dodoi{10.3847/0004-637X/818/2/133}

\bibitem[{{Sing} {et~al.}(2016){Sing}, {Fortney}, {Nikolov}, {Wakeford}, {Kataria}, {Evans}, {Aigrain}, {Ballester}, {Burrows}, {Deming}, {D{\'e}sert}, {Gibson}, {Henry}, {Huitson}, {Knutson}, {Lecavelier Des Etangs}, {Pont}, {Showman}, {Vidal-Madjar}, {Williamson}, \& {Wilson}}]{Sing16}
{Sing}, D.~K., {Fortney}, J.~J., {Nikolov}, N., {et~al.} 2016, \nat, 529, 59, \dodoi{10.1038/nature16068}

\bibitem[{{Stock} {et~al.}(2022){Stock}, {Kitzmann}, \& {Patzer}}]{Stock22}
{Stock}, J.~W., {Kitzmann}, D., \& {Patzer}, A. B.~C. 2022, \mnras, 517, 4070, \dodoi{10.1093/mnras/stac2623}

\bibitem[{Stock {et~al.}(2018)Stock, Kitzmann, Patzer, \& Sedlmayr}]{Stock18}
Stock, J.~W., Kitzmann, D., Patzer, A. B.~C., \& Sedlmayr, E. 2018, Monthly Notices of the Royal Astronomical Society, \dodoi{10.1093/mnras/sty1531}

\bibitem[{Tazaki \& Tanaka(2018)}]{Tazaki18}
Tazaki, R., \& Tanaka, H. 2018, The Astrophysical Journal, 860, 79, \dodoi{10.3847/1538-4357/aac32d}

\bibitem[{Tennyson {et~al.}(2016)Tennyson, Yurchenko, Al-Refaie, Barton, Chubb, Coles, Diamantopoulou, Gorman, Hill, Lam, Lodi, McKemmish, Na, Owens, Polyansky, Rivlin, Sousa-Silva, Underwood, Yachmenev, \& Zak}]{Tennyson16}
Tennyson, J., Yurchenko, S.~N., Al-Refaie, A.~F., {et~al.} 2016, Journal of Molecular Spectroscopy, 327, 73, \dodoi{https://doi.org/10.1016/j.jms.2016.05.002}

\bibitem[{{Tinetti} {et~al.}(2007){Tinetti}, {Vidal-Madjar}, {Liang}, {Beaulieu}, {Yung}, {Carey}, {Barber}, {Tennyson}, {Ribas}, {Allard}, {Ballester}, {Sing}, \& {Selsis}}]{Tinetti07}
{Tinetti}, G., {Vidal-Madjar}, A., {Liang}, M.-C., {et~al.} 2007, \nat, 448, 169, \dodoi{10.1038/nature06002}

\bibitem[{{Tinetti} {et~al.}(2018){Tinetti}, {Drossart}, {Eccleston}, {Hartogh}, {Heske}, {Leconte}, {Micela}, {Ollivier}, {Pilbratt}, {Puig}, {Turrini}, {Vandenbussche}, {Wolkenberg}, {Beaulieu}, {Buchave}, {Ferus}, {Griffin}, {Guedel}, {Justtanont}, {Lagage}, {Machado}, {Malaguti}, {Min}, {N{\o}rgaard-Nielsen}, {Rataj}, {Ray}, {Ribas}, {Swain}, {Szabo}, {Werner}, {Barstow}, {Burleigh}, {Cho}, {Coud{\'e} du Foresto}, {Coustenis}, {Decin}, {Encrenaz}, {Galand}, {Gillon}, {Helled}, {Morales}, {Garc{\'\i}a Mu{\~n}oz}, {Moneti}, {Pagano}, {Pascale}, {Piccioni}, {Pinfield}, {Sarkar}, {Selsis}, {Tennyson}, {Triaud}, {Venot}, {Waldmann}, {Waltham}, {Wright}, {Amiaux}, {Augu{\`e}res}, {Berth{\'e}}, {Bezawada}, {Bishop}, {Bowles}, {Coffey}, {Colom{\'e}}, {Crook}, {Crouzet}, {Da Peppo}, {Sanz}, {Focardi}, {Frericks}, {Hunt}, {Kohley}, {Middleton}, {Morgante}, {Ottensamer}, {Pace}, {Pearson}, {Stamper}, {Symonds}, {Rengel}, {Renotte}, {Ade}, {Affer}, {Alard}, {Allard}, {Altieri}, {Andr{\'e}}, {Arena}, {Argyriou},
  {Aylward}, {Baccani}, {Bakos}, {Banaszkiewicz}, {Barlow}, {Batista}, {Bellucci}, {Benatti}, {Bernardi}, {B{\'e}zard}, {Blecka}, {Bolmont}, {Bonfond}, {Bonito}, {Bonomo}, {Brucato}, {Brun}, {Bryson}, {Bujwan}, {Casewell}, {Charnay}, {Pestellini}, {Chen}, {Ciaravella}, {Claudi}, {Cl{\'e}dassou}, {Damasso}, {Damiano}, {Danielski}, {Deroo}, {Di Giorgio}, {Dominik}, {Doublier}, {Doyle}, {Doyon}, {Drummond}, {Duong}, {Eales}, {Edwards}, {Farina}, {Flaccomio}, {Fletcher}, {Forget}, {Fossey}, {Fr{\"a}nz}, {Fujii}, {Garc{\'\i}a-Piquer}, {Gear}, {Geoffray}, {G{\'e}rard}, {Gesa}, {Gomez}, {Graczyk}, {Griffith}, {Grodent}, {Guarcello}, {Gustin}, {Hamano}, {Hargrave}, {Hello}, {Heng}, {Herrero}, {Hornstrup}, {Hubert}, {Ida}, {Ikoma}, {Iro}, {Irwin}, {Jarchow}, {Jaubert}, {Jones}, {Julien}, {Kameda}, {Kerschbaum}, {Kervella}, {Koskinen}, {Krijger}, {Krupp}, {Lafarga}, {Landini}, {Lellouch}, {Leto}, {Luntzer}, {Rank-L{\"u}ftinger}, {Maggio}, {Maldonado}, {Maillard}, {Mall}, {Marquette}, {Mathis}, {Maxted}, {Matsuo},
  {Medvedev}, {Miguel}, {Minier}, {Morello}, {Mura}, {Narita}, {Nascimbeni}, {Nguyen Tong}, {Noce}, {Oliva}, {Palle}, {Palmer}, {Pancrazzi}, {Papageorgiou}, {Parmentier}, {Perger}, {Petralia}, {Pezzuto}, {Pierrehumbert}, \& {Pillitteri}}]{Tinetti18}
{Tinetti}, G., {Drossart}, P., {Eccleston}, P., {et~al.} 2018, Experimental Astronomy, 46, 135, \dodoi{10.1007/s10686-018-9598-x}

\bibitem[{{Tsai} {et~al.}(2023){Tsai}, {Lee}, {Powell}, {Gao}, {Zhang}, {Moses}, {H{\'e}brard}, {Venot}, {Parmentier}, {Jordan}, {Hu}, {Alam}, {Alderson}, {Batalha}, {Bean}, {Benneke}, {Bierson}, {Brady}, {Carone}, {Carter}, {Chubb}, {Inglis}, {Leconte}, {Line}, {L{\'o}pez-Morales}, {Miguel}, {Molaverdikhani}, {Rustamkulov}, {Sing}, {Stevenson}, {Wakeford}, {Yang}, {Aggarwal}, {Baeyens}, {Barat}, {de Val-Borro}, {Daylan}, {Fortney}, {France}, {Goyal}, {Grant}, {Kirk}, {Kreidberg}, {Louca}, {Moran}, {Mukherjee}, {Nasedkin}, {Ohno}, {Rackham}, {Redfield}, {Taylor}, {Tremblin}, {Visscher}, {Wallack}, {Welbanks}, {Youngblood}, {Ahrer}, {Batalha}, {Behr}, {Berta-Thompson}, {Blecic}, {Casewell}, {Crossfield}, {Crouzet}, {Cubillos}, {Decin}, {D{\'e}sert}, {Feinstein}, {Gibson}, {Harrington}, {Heng}, {Henning}, {Kempton}, {Krick}, {Lagage}, {Lendl}, {Lothringer}, {Mansfield}, {Mayne}, {Mikal-Evans}, {Palle}, {Schlawin}, {Shorttle}, {Wheatley}, \& {Yurchenko}}]{Tsai23}
{Tsai}, S.-M., {Lee}, E. K.~H., {Powell}, D., {et~al.} 2023, \nat, 617, 483, \dodoi{10.1038/s41586-023-05902-2}

\bibitem[{Vahidinia {et~al.}(2024)Vahidinia, Moran, Marley, \& Cuzzi}]{Vahidinia24}
Vahidinia, S., Moran, S.~E., Marley, M.~S., \& Cuzzi, J.~N. 2024, Aggregate Cloud Particle Effects in Exoplanet Atmospheres.
\newblock \doarXiv{2408.11215}

\bibitem[{Visscher {et~al.}(2010)Visscher, Lodders, \& Fegley}]{Visscher10}
Visscher, C., Lodders, K., \& Fegley, B. 2010, The Astrophysical Journal, 716, 1060–1075, \dodoi{10.1088/0004-637x/716/2/1060}

\bibitem[{{Voyer, M.} \& {Changeat, Q.}(2026)}]{Voyer26}
{Voyer, M.}, \& {Changeat, Q.} 2026, \aap, 707, A127, \dodoi{10.1051/0004-6361/202558469}

\bibitem[{{Waterman}(1971)}]{Waterman71}
{Waterman}, P.~C. 1971, \prd, 3, 825, \dodoi{10.1103/PhysRevD.3.825}

\bibitem[{{Wiscombe}(1980)}]{Wiscombe80}
{Wiscombe}, W.~J. 1980, Applied Optics, 19, 1505, \dodoi{10.1364/AO.19.001505}

\bibitem[{Wogan {et~al.}(2024)Wogan, Batalha, Zahnle, Krissansen-Totton, Tsai, \& Hu}]{Wogan24}
Wogan, N.~F., Batalha, N.~E., Zahnle, K., {et~al.} 2024, JWST observations of K2-18b can be explained by a gas-rich mini-Neptune with no habitable surface.
\newblock \doarXiv{2401.11082}

\bibitem[{{Yurkin} \& {Hoekstra}(2011)}]{Yurkin2011}
{Yurkin}, M.~A., \& {Hoekstra}, A.~G. 2011, \jqsrt, 112, 2234, \dodoi{10.1016/j.jqsrt.2011.01.031}

\bibitem[{{Zeidler} {et~al.}(2011){Zeidler}, {Posch}, {Mutschke}, {Richter}, \& {Wehrhan}}]{Zeidler11}
{Zeidler}, S., {Posch}, T., {Mutschke}, H., {Richter}, H., \& {Wehrhan}, O. 2011, \aap, 526, A68, \dodoi{10.1051/0004-6361/201015219}

\end{thebibliography}
\bibliographystyle{aasjournal}

\end{document}